\documentclass[
]{ceurart}

\usepackage[]{todonotes}

\usepackage{listings}
\newcommand{\npm}{\textsf{npm}\xspace}
\newcommand{\cran}{\textsf{CRAN}\xspace}
\newcommand{\cpan}{\textsf{CPAN}\xspace}
\newcommand{\packagist}{\textsf{Packagist}\xspace}
\newcommand{\rubygems}{\textsf{RubyGems}\xspace}
\newcommand{\cargo}{\textsf{Cargo}\xspace}
\newcommand{\nuget}{\textsf{NuGet}\xspace}
\newcommand{\maven}{\textsf{Maven}\xspace}
\newcommand{\pypi}{\textsf{PyPI}\xspace}
\newcommand{\github}{\textsf{GitHub}\xspace}
\newcommand{\gha}{\textsf{GitHub Actions}\xspace}
\newcommand{\gitlab}{\textsf{GitLab}\xspace}
\newcommand{\js}{JS\xspace}
\newcommand{\dependabot}{\textsf{Dependabot}\xspace}

\begin{document}

\copyrightyear{2024}
\copyrightclause{Copyright for this paper by its authors.
  Use permitted under Creative Commons License Attribution 4.0
  International (CC BY 4.0).}

\conference{Accepted for BENEVOL24, the 23rd Belgium-Netherlands Software Evolution Workshop}

\title{An Overview and Catalogue of Dependency Challenges\\
in Open Source Software Package Registries}


\author[1]{Tom Mens}[%
orcid=0000-0003-3636-5020,
email=tom.mens@umons.ac.be
]
\author[1,2]{Alexandre Decan}[%
orcid=0000-0002-5824-5823,
email=alexandre.decan@umons.ac.be
]

\address[1]{Software Engineering Lab, University of Mons, Belgium}
\address[2]{FRS-FNRS Research Associate}


\begin{abstract}
While open-source software has enabled significant levels of reuse to speed up software development, it has also given rise to the dreadful dependency hell that all software practitioners face on a regular basis.
This article provides a catalogue of dependency-related challenges that come with relying on OSS packages or libraries. The catalogue is based on the scientific literature on empirical research that has been conducted to understand, quantify and overcome these challenges.
Our overview of this very active research field of package dependency management can be used as a starting point for junior and senior researchers as well as practitioners that would like to learn more about research advances in dealing with the challenges that come with the dependency networks of large OSS package registries.
\end{abstract}


\maketitle

\section{Introduction}

Probably every complex software system today relies, to some extent, on reusable OSS libraries distributed through package managers hosting millions of libraries in their package registries. Such reuse inevitably leads to what is commonly known to developers as the \emph{dependency hell}. Software becomes dysfunctional, outdated, buggy, or insecure due to package interdependencies and updates that lead to conflicts, breaking changes, incompatibilities, security issues, deprecations, and many more. Dealing with such issues requires investing significant time and effort. This is why a lot of empirical research in the last decade has focused on understanding OSS package dependency networks, and on mechanisms to cope with dependency-related challenges.

We provide an overview of the  literature on how OSS package reuse practices have evolved in recent years.
We propose a catalogue of challenges in OSS package dependency networks and beyond, and present recent empirical research to understand and address each of these challenges.
It can be used as a basis for junior and senior researchers as well as practitioners that would like to get a kick-start in the state-of-the-art research challenges and advances in package dependency management.

\section{Starting point of the overview}
\label{sec:RQ}

As a starting point for this article we based ourselves on a seminal article reporting on an empirical comparison  of the evolution of the dependency networks of seven of the largest package registries for mainstream programming languages~\cite{Decan2019EMSE}.
%
The article provided quantitative insights into the dependency issues that software practitioners face when relying on reusable OSS libraries distributed through large package registries. 
The article focused on package registries for mainstream programming languages for which complete package metadata was available from the \textsf{libraries.io} monitoring service, including reliable information about package dependencies.
Seven ecosystems were 
studied over a five-year  period (2012 -- 2016): \cargo for Rust, \cpan for Perl, \cran for R, \npm for JavaScript (\js), \nuget for the .NET platform, \packagist for PHP, and \rubygems for Ruby.
The  study answered four main research questions pertaining to the evolution of the dependency networks of these  package registries:

\textbf{RQ1: How do package dependency networks grow over time?}
The dependency networks of all studied package registries were observed  to grow over time, though the speed of growth differs, with \npm being the largest registry experiencing the fastest growth. 
The dependency network's complexity in terms of ratio of dependencies over packages was observed to remain stable for \cpan, \packagist and \rubygems, while it tended to increase for \cargo, \cran, \npm and \nuget.

\textbf{RQ2: How frequently are packages updated? }
A \emph{Changeability Index} was defined to characterise a registry's propensity to change at time $t$. 
The number of package updates was observed to remain stable for  \cargo, \cpan and \cran, while it had a tendency to grow for the other registries.
Most package releases were observed to receive updates within a few months. However, the number of package updates was not evenly distributed across packages, with a minority of active packages responsible for most of the package updates. Younger or required packages were found to receive package updates more often. Some of the observed behaviours depended on the age of the package registry.

\textbf{RQ3: To which extent do packages depend on other packages?}
A  \emph{Reusability Index} was defined to measure the amplitude of reuse (number of required packages) and the extent of reuse (number of dependent packages) at time $t$. 
Dependencies were observed to abound in all package registries. Most packages depend on other packages, and the proportion of connected packages increases over time. Dependencies were not evenly spread across packages: $<30\%$ of the packages were required by other packages, and $< 17\%$ of all required packages concentrated $> 80\%$ of all reverse dependencies. This unequal concentration increased over time.

\textbf{RQ4: How prevalent are transitive dependencies?}
The indirect reuse induced by the prevalence of transitive dependencies in a package dependency network causes package failures to propagate. This may impact large parts of the network.
The majority of dependent packages were observed to have few direct dependencies but many transitive dependencies. More than half of the top-level packages have a dependency tree of depth 3 or higher.
The \emph{$p$-Impact Index} of a package registry at time $t$ was defined to
quantify the number of packages that could have a high potential impact because of their many transitive dependents. 
A  notable increase in this impact over time was observed for \cargo, \npm and \nuget, suggesting that these registries are becoming more subject to single points of failure.

\medskip

\begin{table}[!t]
   \centering
   \caption{Comparing the number of packages in seven package registries between April 2017 and October 2024.}



  \label{tab:secosize}
{\small
   \begin{tabular}{ccrrr}
      \toprule
       Package    & Language & \multicolumn{2}{c}{\# packages} & Increase\\
       manager & & 04-2017 & 10-2024 & factor\\
      \midrule
      \npm       & \js  & 462K & 4,875K & 10.6\\
      \nuget   &  .NET & 84K & 539K & 6.4\\
      \packagist & PHP & 97K & 461K & 4.8\\
      \rubygems & Ruby & 132K & 187K & 1.4\\
      \cargo      & Rust  & 9K & 167K & 18.6 \\
      \cran      & R  & 12K & 27K & 2.3\\
      \cpan      & Perl     & 34K & 41K & 1.2 \\
      \bottomrule
   \end{tabular}
}
\end{table}

Based on the answers to these RQs, the authors observed that three package registries (\npm, \nuget and \cargo) faced more difficulties to cope with the rapid growth and increasing complexity of their dependency networks, suggesting that they should make an effort to reduce their complexity and fragility.
Table~\ref{tab:secosize} (based on data obtained from \textsf{libraries.io}) shows that these registries have continued their growth, suggesting that this observation still holds today.
Comparing the number of packages in the dataset of \cite{Decan2019EMSE} in April 2017 with the data in October 2024, we observe an 18-fold increase in the number of packages for \cargo, a 10-fold increase for \npm, and a 6-fold increase for \nuget.

\smallskip
Due to a lack of complete or reliable dependency data, only seven package registries were included in \cite{Decan2019EMSE}.
\citet{Dietrich2019MSR} considered a larger collection of 17 different package managers, investigating over 70 million dependencies, complemented by a survey of 170 developers.
Similar in vein, \citet{Bogart2021TOSEM} combined a survey, repository mining, and document analysis to observe the dependency practices across 18 ecosystems and their communities. They observed that all ecosystems share values such as stability and compatibility, but differ in other values, and use different tools, policies and practices to support these values. This implies that findings for one ecosystem may not generalise to another.
Researchers have therefore empirically studied dependency issues in specific package registries, such as \maven for Java~\cite{Soto-Valero2021,Ochoa2022EMSE,Reyes2024SANER}, \pypi for Python~\cite{Valiev2018PyPI,Alfadel2023EMSE,Cao2023TSE}, Swift~PM for Swift~\cite{Rahkema2023}, the ROS ecosystem~\cite{Kolak2020ICSME}, and \cran \cite{Mora-Cantallops2020JSS,Mora-Cantallops2020JSEP}.

\section{Dependency-related challenges}
\label{sect:challenges}

The empirical findings of \cite{Decan2019EMSE} revealed the increasing (transitive) complexity, impact, and growth of OSS package registries. This makes it challenging for developers to maintain (dependencies on) such packages.
The remainder of this article presents a literature review on empirical research focusing on these challenges and on strategies and solutions that have been proposed to overcome them.
The review is based on the content of scientific articles that appeared since 2018, 
the publication year of \cite{Decan2019EMSE}. 
 We considered articles in major software engineering conferences or journals that directly cited this work, and used snowballing to include more recent relevant articles. To avoid missing out on important  research advances, we also 
searched through Google Scholar, Semantic Scholar and ResearchRabbit to identify other relevant recent empirical research in this domain.
Table~\ref{tab:dep-issues} catalogs the dependency-related challenges that we have been able to identify based on our literature review.
The remainder of this section discusses the most relevant recent research for each of these challenges.

\begin{table}[!t]
   \centering
   \caption{Catalogue of dependency-related challenges and associated scientific references since 2018. Additional references are provided in the respective subsections of Section~\ref{sect:challenges}.}
   \label{tab:dep-issues}
   {\small
   \begin{tabular}{p{7cm}p{7cm}}
      \toprule
       Dependency-related challenges    & References\\
      \midrule
      Outdated dependencies & \cite{Kula2018EMSE,Decan2018ICSME,Zerouali2018,Lauinger2018,Zerouali2019JSEP,Stringer2020APSEC,Zerouali2021SCP,Hejderup2022JSS,Zerouali2023MSR,Decan2023JSS}\\
      Breaking changes and backward incompatibilities& \cite{Bogart2021TOSEM,Mezzetti2018,Moller2019ESECFSE,Mujahid2020MSR,Brito2020EMSE,Zhang2022ASE,Venturini2023}\\
      Versioning policies and update strategies &\cite{Dietrich2019MSR,Ochoa2022EMSE,Lam2020semver,Cogo2021TSE,Decan2021TSE,Decan2021SCP,Decan2022TSE,JavanJafari2023TOSEM} \\
      Dependency solving &  \cite{Mancinelli2006,Abate2012JSS, Abate2020SANER, Pinckney2023ICSE} \\
      Bloated and missing dependencies  & \cite{Soto-Valero2021,Cao2023TSE,Jafari2022TSE,SotoValero2023TOSEM,Weeraddana2024}\\
      Vulnerable dependencies & \cite{Alfadel2023EMSE,Lauinger2018,Chinthanet2021EMSE,Decan2018MSR, Prana2021, Zerouali2022, Liu2022ICSE, Zimmermann2019, Alfadel2021MSR, Mohayeji2023, Alfadel2023TOSEM, Wang2023TSE}\\
      Supply chain attacks & \cite{Ohm2020,Ohm2020DIMVA,Duan2021NDSS,Enck2022,Lamb2022,Wermke2023,Fourne2023,ODonoghue2024} \\
      Library deprecation and migration & \cite{Cogo2022TSE,He2021ESEC,He2021SANER,Gu2023SANER,Mujahid2022ASE}\\
      Depending on trivial libraries & \cite{Abdalkareem2020EMSE,Chen2021EMSE,Chowdhury2021TSE}\\
      Abandoned and unmaintained dependencies & \cite{Avelino2019empirical,Kaur2022JSEP,Miller2023,Zimmermann2023EMSE} \\
      Incompatible licenses & \cite{Xu2023TOSEM,Xu2023ISSTA,XuLicense2023,Wu2024MSR} \\
      \bottomrule
   \end{tabular}
   }
\end{table}

\subsection{Outdated dependencies}
\label{sec:outdatedness}

Updating one's dependencies is a good strategy to reduce or avoid many dependency issues. By keeping dependencies up to date, one can benefit from the most recent functionalities, bug fixes and vulnerability fixes. Staying up to date also makes it easier to interact with upstream dependency providers, who tend to focus on their most recent package releases.
\citet{Kula2018EMSE} studied 4,600 \github software projects and 2,700 library dependencies, revealing that 81.5\% of the studied systems have outdated dependencies.
To quantify such outdatedness of a package w.r.t. its dependencies, the \emph{technical lag} concept has been introduced 
\cite{Decan2018ICSME,Zerouali2018,Stringer2020APSEC}. Different from the concept of \emph{technical debt}, which focuses on the internal code quality of a software system, technical lag quantifies how much a package is ``lagging behind'' w.r.t upstream --often third-party-- dependencies. Such lag can be expressed along different dimensions~\cite{Zerouali2019JSEP}: \emph{time lag} (e.g., the time interval between the current version of a dependency being used and some more recent version); \emph{version lag} (how many major/minor/patch versions a dependency is behind); \emph{security lag} and \emph{bug lag} (if more recent versions have known fixes for vulnerabilities or bugs that affect the version being used). Which dimension or combination to use ultimately remains the decision of the package maintainer, depending on one's priorities.

Researchers have extended the analysis of outdatedness beyond the boundaries of package registries.
\citet{Lauinger2018} analysed the reliance of 133K websites on \js
libraries, observing that a majority of these websites are at least one patch version behind for one of their included libraries, and that most of them are relying on library versions that are outdated by several years.
\citet{Zerouali2021SCP} studied outdatedness in Docker, the most popular containerization technology.
Considering over 3K container images in Docker Hub, they empirically quantified their outdatedness w.r.t. installed \js, Python and Ruby packages.
 %
\citet{Zerouali2023MSR}  studied the outdatedness of 9,482 Helm charts, configuration files for containerized applications for Kubernetes distributed through the Helm package manager.
They observed that around half of the container images used in Helm charts are outdated and nearly nine out of them  are exposed to vulnerabilities.
%
\citet{Decan2023JSS} studied the reliance of \gha automation workflows on reusable Actions. They found that these reusable Actions are frequently updated, and that most of the workflows are relying on outdated Action versions, hence lagging behind the latest available version for at least seven months, even though they had the opportunity to be updated during at least nine months.

\emph{Tool support.} Given the importance of keeping dependencies up to date, many automated tools emerged to help developers in this task, such as \dependabot (now part of \github), \textsf{Gemnasium} (now part of \gitlab), the independent multi-platform solution \textsf{Renovate}, and \textsf{Greenkeeper} (no longer available).
Based on a mixed method empirical analysis, \citet{He2023TSE} evaluated the effectiveness of \dependabot in keeping dependencies up to date, observing that projects reduce their technical lag after its adoption. On the downside, \dependabot was found to recommend too many incompatible updates; and the amount of \dependabot notifications was considered too high.
\citet{Rombaut2023TOSEM} analysed 93,196 issues opened by \textsf{Greenkeeper} for \npm projects hosted on \github. \textsf{Greenkeeper} was found to induce a significant amount of overhead and false alarms in reported issues.
\citet{Hejderup2022JSS} recommended improving existing dependency update tools to combine static and dynamic analysis in order to reduce the number of semantically conflicting updates.
\citet{Dann2023ICSE} proposed UPCY, an automated dependency update tool that aims to minimise incompatible dependencies when updating.
The tool was validated on 29,698 updates in 380 \maven projects, observing an important improvement compared to the updates recommended by existing tools.

\subsection{Breaking changes and backward incompatibilities}

Keeping dependencies up to date requires much effort from developers, because of the rapid pace of package updates, but also because these updates can lead to \emph{backward incompatibilities} due to the introduction of \emph{breaking changes}.
\citet{Bogart2021TOSEM} investigated the policies and practices of making and facing breaking changes in 18 software ecosystems. They observed that maintainers are frequently exposed to breaking changes, and that ecosystems differ in their approaches to breaking changes.
Through a mixed methods empirical study \citet{Brito2020EMSE}  analysed  why and how developers introduce breaking changes in libraries. The identified reasons were to support new features, simplify existing APIs, and improve maintainability. They also identified a contrast between library producers and consumers in the perceived effort to overcome breaking changes. On the one side, according to the developers, the effort to adopt these breaking changes is generally limited. On the other side, nearly half of the questions related to breaking changes on StackOverflow are about how to integrate and overcome these breaking changes.

\citet{Venturini2023} studied \emph{backward incompatibilities} introduced by breaking changes when upgrading \npm dependencies to newer versions. By analysing dependency updates in 384 \npm packages they found that 11.7\% of them lead to  breaking changes, even though 44\% of these dependency upgrades  were meant to be backward compatible. 
They observed that more than half of the backward incompatible updates are due to transitive dependencies, and that the usual mitigation strategy is either for the provider package to release a patch fixing the backward incompatibilities (usually within a week) or for the dependent packages to incorporate these backward incompatible changes in newer version of their packages (taking 4 months on average).
In a similar vein, \citet{Jayasuriya2023ISSTA} studied the \maven ecosystem by analysing 142K+ 
direct dependencies of 18K+ 
 \maven artifacts. 71.6\% of the dependencies were outdated, and 11.6\% of the dependency upgrades applied to them resulted in breaking changes. 
Changes in transitive dependencies were a major factor for these breaking changes.
%

Recently, researchers have started to focus on so-called \emph{semantic} or \emph{behavioural breaking changes}. 
\citet{Jayasuriya2024ACM} conducted an empirical analysis on 30,548 dependencies of 8,086 \maven artifacts
 to identify the impact of dependency upgrades on behavioral breaking changes in the test suites of client Java projects.
Only 2.30\% of the dependency upgrades caused client tests to break.
\citet{Zhang2022ASE} proposed and empirically validated a tool to statically detect semantic breaking changes in third-party libraries used by Java projects  by measuring semantic differences.

\emph{Tool support.} Several language-specific tools help developers to detect breaking changes before an update is released to the clients.
Examples include \textsf{PyCompat}~\cite{Zhang2020SANER}, \textsf{DepOwl}~\cite{Jia2021ICSE} and \textsf{AexPy}~\cite{Du2022ISSRE} for Python, \textsf{APIDiff}~\cite{Brito2018APIDiff}, \textsf{Clirr} and \textsf{Revapi} for Java~\cite{jezek2017api}, and \textsf{NoRegrets} for \js~\cite{Mezzetti2018,Moller2019ESECFSE}. 
These tools analyse whether the changes made to the types used in the public API may break clients.
Empirical evidence revealed that such tools are able to catch most breaking changes in practice \cite{Dig2006}.
Approaches based on type regression testing \cite{Mezzetti2018,Mujahid2020MSR} run the test suites of a library’s clients to detect breaking changes. While effective, it can require a lot of storage capacity and execution time.
\citet{Moller2019ESECFSE} propose an improved variant of type regression testing for \js libraries, by automatically generating tests from a reusable API model. This approach is shown to run faster and find breaking changes in more libraries.

\subsection{Versioning policies and update strategies}

Proper versioning policies and update strategies, such as \emph{semantic versioning}~\cite{Dietrich2019MSR,Lam2020semver,JavanJafari2023TOSEM} can help in mitigating breaking changes. Semantic versioning provides an implicit convention between the library consumers (who specify the range of allowed versions) and library producers (who avoid introducing breaking changes in non-major versions).
\citet{Decan2021TSE} investigated semantic versioning compliance in \cargo, \npm, \packagist and \rubygems, observing that most packages are compliant with semantic versioning.
In a follow-up work~\cite{Decan2021SCP}, they found that semantic versioning is misused in packages not having reached the 1.0.0 version barrier.
By analysing 120K library upgrades on \maven, \citet{Ochoa2022EMSE}  found that only a minority of library consumers are affected by breaking changes, and a large majority of the upgrades comply with semantic versioning.
A study by \citet{Jayasuriya2023ISSTA} on dependency upgrades of outdated \maven libraries leading to breaking changes revealed that almost half of these breaking changes coincided with violations of the semantic versioning scheme. 

While convenient to inform about the presence of breaking changes, semantic versioning does not help developers in overcoming them, and important updates containing bug or security fixes may still be missed due to the breaking changes that come with them.
\emph{Backporting} these fixes from more recent releases to less recent ones may be the only solution to benefit from them.
\citet{Decan2022TSE} empirically studied such backported changes in \cargo, \npm, \packagist and \rubygems. They found infrequent use of backporting to maintain previous major versions, even when those versions were still widely used. The lack of backports led thousands of packages exposed to security vulnerabilities even if a fix was available for them.
Similar in vein,  \citet{Cogo2021TSE} investigated dependency \emph{downgrades} in \npm. By analysing release notes and commit messages, they found that maintainers downgrade their dependencies either reactively (to avoid defects in a specific version, or to cope with unexpected feature changes and incompatibilities) or pro-actively (to avoid issues in future releases). Moreover, maintainers tend to be more conservative on the dependency versions they use after such downgrades.

\subsection{Dependency solving}
\label{sec:depsolving}

One of the key responsibilities of package managers is \emph{dependency solving}, in order to ensure that all versions of all installed packages are mutually compatible and non-conflicting, and remain so when removing and upgrading existing packages or installing new ones \cite{Abate2020SANER}. Unfortunately, dependency solving has shown to be an NP-complete problem for most package managers \cite{Mancinelli2006,Abate2012JSS}. 
Many package managers provide ad hoc solutions that are not always complete or lack expressiveness.

\citet{Pinckney2023ICSE} proposed and evaluated Maxnpm and Pacsolve to overcome the shortcomings of the npm dependency solver. It   
enables customizable constraints and optimization goals, empowering developers to combine multiple objectives when installing dependencies. 
Other researchers have proposed more generic formally founded solutions based on constraint solving and optimisation \cite{Mancinelli2006}.

\emph{Functional package managers}, such as GNU Guix\footnote{\url{https://guix.gnu.org}} and Nix\footnote{\url{https://nixos.org}}, provide a robust solution by 
enforcing a declarative approach to dependency management, requiring all dependencies to be declared upfront to build or run software. This ensures that the software environment is fully defined, with each dependency explicitly specified. Additionally, these managers  enable the creation of separate namespaces on-the-fly, allowing multiple versions of the same package to be installed side-by-side without any risk of incompatibility or inconsistencies. 
This represents a significant improvement over mainstream package managers, which often struggle to achieve similar functionality without complex workarounds or dependency resolution issues.

\subsection{Bloated and missing dependencies}

Several researchers have studied dependency smells related to \emph{bloated} and \emph{missing} dependencies, for the dependency networks of \maven \cite{Soto-Valero2021}, \pypi \cite{Cao2023TSE} and \npm \cite{Jafari2022TSE}.
Missing dependencies refer to required packages that are not explicitly declared in the configuration file and hence need to be manually installed to avoid dependency problems.
Bloated or unused dependencies are packaged with the application’s compiled code but are actually not necessary to build and run the application. Including them can
increase the size of the application and possibly affect its performance and security posture.
\citet{SotoValero2023TOSEM} proposed a novel technique for debloating dependencies in Java projects. The technique relies on bytecode coverage analysis to precisely capture what parts of a project and its dependencies are used when running with a specific workload. 68\% of the bytecode of Java libraries and 20\% of their total dependencies could be removed through coverage-based debloating. Based on a dataset of 988 client projects, 81\% of them successfully compiled and passed their test suite when the original library was replaced by its debloated version.
\citet{Weeraddana2024} showed that unused dependencies also negatively impact CI resource usage. Based on 20K+ commits in 1,487 projects relying on \npm packages, they found that $>55\%$ of the CI build time was associated with dependency updates triggered by unused dependencies.

\subsection{Vulnerable dependencies}

A very important challenge for OSS package registries is how to cope with vulnerabilities and security weaknesses in dependencies, either directly or indirectly.
They can lead to single points of failure with a huge cascading impact through transitive dependencies. One example of a major incident was discovered in May 2021, with a remote code execution vulnerability in \npm's \textsf{pac-resolver}  package that received over 3 million weekly downloads \cite{pacresolver}. 
Another example in December 2021 was a vulnerability in \maven's \textsf{Log4Shell} package in the \textsf{Log4j} logging framework for Java that caused widespread damage \cite{Hiesgen2022}. These incidents increased public awareness of the need to mitigate OSS supply chain attacks~\cite{Duan2021NDSS}.

Because of their importance and impact, security vulnerabilities in package registries are a very active domain of research.
Dependency outdatedness is one of the major sources of security vulnerabilities.
Starting from a dataset of 133K websites depending on \js libraries, \citet{Lauinger2018} observed that 37\% of the outdated websites included at least one vulnerable library.
\citet{Decan2018MSR} studied vulnerabilities in \npm packages and observed that while vulnerabilities are quickly fixed after their discovery, it takes a lot of time  to adopt the fix for a large fraction of packages that (transitively) depend on the vulnerable package. The main reasons are too restrictive dependency constraints and unmaintained packages. \citet{Zimmermann2019} aligned with these insights, observing that a lack of maintenance by a small number of maintainers causes many \npm packages to depend on vulnerable, unmaintained packages. This confirms that \npm suffers from single points of failure.
\citet{Chinthanet2021EMSE} analysed the lag between the vulnerable release and its package-side fixing release. Through an empirical study of the adoption and propagation tendencies of 1,290 package-side fixing releases that impact throughout a network of 1.5M+ releases of \npm packages, they found that stale clients require additional migration effort, even if the package-side fixing release was quickly made available.
\citet{Liu2022ICSE} investigated the security threats from vulnerabilities in the \npm dependency network. They  constructed a dependency-vulnerability knowledge graph capturing 10M+  library versions and 60M+ dependency relations and proposed an algorithm to statically resolve dependency trees and transitive vulnerability propagation paths. Based on it, they empirically studied the evolution of vulnerability propagation in \npm. Among many findings, they confirmed outdatedness to be a major source of vulnerable dependencies. To cope with it, they developed \textsf{DTreme}, a vulnerability remediation method that outperforms the official \textsf{npm audit fix} tool.
\citet{Prana2021} analysed vulnerabilities in libraries used by 450 projects written in Java, Python, and Ruby. They examined the types, distribution, severity and persistence of the vulnerabilities over a one-year period. They found that most vulnerabilities persist throughout the observation period, and the resolved ones take 3-5 months to be fixed.

Several studies have compared \npm to other package registries, given that the particularities of specific ecosystems could affect their vulnerability posture.
\citet{Zerouali2022} compared \npm to \rubygems w.r.t. how and when vulnerabilities are disclosed and fixed, how their prevalence changes over time, and how vulnerable packages expose their (transitive) dependents to vulnerabilities.
Among many findings, they observed an increase in vulnerabilities in \npm, but also a faster disclosure for \rubygems. Moreover, vulnerabilities in \npm tend to affect fewer package releases.
\citet{Alfadel2023EMSE} found similar vulnerability characteristics in \pypi and \npm, as well as divergences that could be attributed to specific \pypi policies. They empirically studied 1,396 vulnerability reports affecting 698 \pypi packages. Focusing on 2,224 Python projects they observed that more vulnerabilities are discovered over time, and a large portion ($>40\%$) are only fixed after having been publicly announced. Moreover, more than half of the dependent projects rely on at least one vulnerable package, and it requires seven months to update to a non-vulnerable version.

\emph{Tool support.} Many tools are available to help developers detect and resolve security weaknesses in vulnerable dependencies. One of them is \github's \dependabot, which issues pull requests to automatically update vulnerable dependencies, providing an effective platform for increasing awareness of dependency vulnerabilities and mitigating vulnerability threats.
\citet{Alfadel2021MSR} studied the degree to which developers adopt \dependabot by investigating 2,904 OSS \js projects. They observed that a majority of security-related pull requests are accepted, often merged within a day.
The severity of the dependency vulnerability and the potential risk of breaking changes were not strongly correlated with the time to merge these pull requests.
\citet{Mohayeji2023} empirically studied receptivity to \dependabot security updates in \js projects, observing that developers tend to delegate the task of fixing vulnerable dependencies and merge the majority of recommended security updates within several days. This is considerably faster than fixing vulnerabilities manually, which often takes up to several months.
Similar to earlier research, \citet{Alfadel2023TOSEM} empirically analysed vulnerable dependencies in 6,546 \js applications, observing that 4.63\% of them were exposed to dependencies with publicly known vulnerabilities, even if a fix was available in 90.8\% of the cases. They proposed \textsf{DepReveal}, a tool to help developers better understand vulnerabilities in their application dependencies and to plan their project maintenance.
\citet{Wang2023TSE} studied 356,283 active \npm packages, observing that, in their latest release, 20\% of them still introduce vulnerabilities via transitive dependencies despite the involved vulnerable packages already fixed the vulnerability for over a year. They empirically studied and distilled the remediation strategies to mitigate the fix propagation lag. Based on this, they developed \textsf{Plumber}, a tool to derive customised remediation suggestions for pivotal packages. The tool received positive feedback from many well-known projects.

So-called \emph{Software Composition Analysis (SCA)} tools are being increasingly adopted by practitioners to keep track of potential security risks due to vulnerable dependencies on third-party libraries. 
Most SCA tools are based on static and/or dynamic analysis of the project dependency graph.
\citet{Imtiaz2021} compared nine industry-leading SCA tools on a case study. They observed important variations in the accuracy of vulnerability reporting, suggesting that practitioners should not rely on any single tool, and that SCA tools need to achieve higher precision by avoiding false positives.
\citet{Dietrich2024} focused on the inverse problem of false negatives (i.e., low recall), when SCA tools miss dependencies on vulnerable components. They found empirical evidence of this for the \maven ecosystem, where somehow obfuscated clones of vulnerable components are deployed in Maven Central, but not marked as vulnerable in vulnerability databases. 

\subsection{Supply chain attacks}

Package dependency networks are part of the wider concept of \emph{software supply chains}~\cite{Duan2021NDSS, Enck2022, Wermke2023}. Assuring their security is crucial to avoid \emph{supply chain attacks} such as the malicious update of the SolarWinds Orion monitoring software, shipping a delayed-activation trojan horse that affected thousands of organisations, including the US government~\cite{Alkhadra2021}.
\citet{Ohm2020DIMVA} analysed 174 malicious software packages in \npm, \pypi, and \rubygems that were used in real-world supply chain attacks.
Nonprofit foundations such as OWASP aim to improve software security in various ways. Related to package dependencies in the software supply chain, they reported in 2022 \emph{Dependency Chain Abuse} to be one of the top 10 CI/CD security risks\footnote{\url{https://owasp.org/www-project-top-10-ci-cd-security-risks/CICD-SEC-03-Dependency-Chain-Abuse}}.
It comes with four main attack vectors: (1) \emph{Dependency confusion} aims to  to trick clients into downloading a malicious package from a public repository rather than a private internal package with the same name; (2) \emph{Dependency hijacking} aims to compromise clients that upgrade their dependency version to a more recent, malicious version; (3) \emph{Typosquatting} aims to mislead clients into using a malicious package that has a very similar name as the intended package; (4) \emph{Brandjacking} aims to mislead clients in depending on malicious packages that have the characteristics of packages of a trusted brand.

To mitigate supply chain attacks, so-called \emph{software bills of materials} (SBOM) have been proposed as complete, formally structured lists of all software components present in a software product, including their licenses, versions, security vulnerabilities, and vendors.\footnote{The most common SBOM formats  are CycloneDX and SPDX.} SBOMs are imposed or recommended by US Executive Order 14028~\cite{ExecutiveOrder} and the EU Cyber Resilience Act~\cite{CRA2022} to facilitate the transparent management of the software supply chain.
\citet{Nocera2023ICSME} observed a low adoption of SBOM in public \github repositories of OSS projects, but with an increasing trend.
Although there are significant efforts from academia and industry to facilitate SBOM development, it is still unclear how practitioners perceive SBOMs and what are the challenges of adopting SBOMs in practice. \citet{ICSE2023Xia} conducted a survey and interviews with SBOM practitioners to understand the current state of SBOM practice, tooling support and concerns for SBOM. They identified several open challenges that need to be further studied, mitigated and addressed.
In a similar vein, \citet{Stalnaker2024ICSE} surveyed 138 practitioners and identified 12 major challenges concerning the creation and use of SBOMs. They propose and discuss four actionable solutions to these challenges, and provide suggestions for future research and development.

\emph{Tool support.} SBOMs enhances vulnerability detection and facilitates license compliance (see Section~\ref{sec:licence}), through the use of SBOM generating tools such as \textsf{Trivy}, \textsf{Syft}, Microsoft's \textsf{sbom-tool}, and GitHub's Dependency Graph. A prerequisite is that such tools achieve full precision and correctness. 
Unfortunately, \citet{YuSBOM2024} observed that the SBOMs generated by these tools were inconsistent and contained dependency omissions, leading to incomplete and perhaps erroneous SBOMs. They consequently proposed  best practices for SBOM generation and introduced a benchmark to steer the development of more robust SBOM generators.
In a similar vein,  \cite{ODonoghue2024} observed a high variability in vulnerability reporting by different SBOM generators.
Tools such as OWASP Dependency-Track\footnote{\url{https://dependencytrack.org}} leverage the capabilities of SBOM to identify and reduce risk in the software supply chain.

\emph{Solutions.} Because secure software supply chains can be hard to attain in practice, 
the Linux Foundation proposes \emph{Supply chain Levels for Software Artifacts (SLSA)}~\footnote{\url{slsa.dev}} 
as a set of guidelines for supply chain security. Higher levels come with increasing security guarantees, with SLSA L3  providing the highest degree of confidence that the generated SBOM is precise, accurate and has not been tampered with. One way to achieve it is by resorting to  the solution of \emph{reproducible builds} \cite{Lamb2022, Fourne2023}.
They ensure that, given the same source code, build environment and build instructions, bitwise identical copies of all  artifacts are created.
The functional package managers mentioned in Section~\ref{sec:depsolving} are instances of this solution. By requiring all dependencies to be declared upfront, they offer a robust solution to dependency management issues, ultimately supporting more reliable software supply chains.

%

\subsection{Library deprecation and migration}

Package registries such as \npm or \cargo allow deprecating package releases, e.g., when a specific release is known to be vulnerable, faulty or incompatible. The solution would be to upgrade or downgrade one's dependencies to that package~\cite{Cogo2021TSE,JavanJafari2023TOSEM}.
Package maintainers could also decide to fully deprecate the package if they decide for some reason to cease maintaining it. In that case, the best possible alternative would be to replace one's dependency on that package with alternative packages. Sticking to the deprecated one is likely to lead to security vulnerabilities for which no fixes will be provided.
~\citet{Cogo2022TSE} quantified deprecation in the \npm registry, observing that 3.7\% of all packages have at least one deprecated release, 31\% of those packages  do not have any replacement release, and 66\% of such packages even deprecated all their releases. They found that 27\% of the client packages directly (and 54\% transitively) depend on at least one deprecated release.

Given that \emph{library migration} is often the only way to deal with deprecated packages, it has received quite some attention from researchers.
Based on an empirical analysis of commits in 19K+ Java projects on GitHub, \citet{He2021ESEC} identified 14 different migration reasons, of which the most important ones were lack of maintenance, known bugs and vulnerabilities, lack of usability, missing features, poor performance, lack of popularity, complexity, integration problems, and licensing issues.
\citet{He2021SANER}  improved upon existing techniques to recommend the most appropriate library to migrate to, based on filtering approaches that leverage the wisdom of the crowd. Candidate libraries are ranked based on a series of metrics. 
In a similar vein, \citet{Mujahid2022ASE} proposed an approach to automatically identify \npm packages requiring replacement, and suggesting alternatives based on wisdom of the crowd. An evaluation showed that 96\% of the suggested alternatives were accurate, and 67\% of surveyed \js developers responded they consider using these suggestions in the future.
\citet{Gu2023SANER} opened up the analysis of library migrations by comparing \maven, \npm and \pypi. Library migrations were found to be prevalent and similar in nature in all three package registries. For \pypi specifically, an increasing competition was observed between libraries.

\subsection{Depending on trivial libraries}

\emph{Trivial} libraries are packages that are either very small or provide little functionality.
Depending on such trivial libraries may unnecessarily introduce a high dependency overhead.
Some trivial libraries have even led to major incidents, as was the case when the \textsf{leftpad} package was removed from \npm in March 2016. This single point of failure caused a breakdown of popular web applications including Facebook and Netflix.
It drove researchers to scrutinise the necessity and prevalence of \emph{trivial packages} in package registries.

In an empirical study of  trivial packages in the \npm and \pypi package registries, \citet{Abdalkareem2020EMSE}  observed that such packages are quite common, making up 16.0\% of \npm and 10.5\% of \pypi.
125 surveyed developers who use trivial packages reported using them because they were \emph{perceived} to be well implemented and tested pieces of code. Contrary to developers' beliefs, only around 28\% of \npm and 49\% of \pypi trivial packages were found to have tests. Surveyed developers were also concerned by the maintenance overhead of depending on trivial packages, which was quantitatively confirmed since 18.4{\%} of the \npm  and 2.9{\%} of the \pypi  trivial packages had more than 20 dependencies.
\citet{Chen2021EMSE} conducted another survey with 59 \js developers who publish trivial \npm packages. The main reported reasons for publishing them were to provide reusable components, testing and documentation, and separation of concerns.  On the downside, the surveyed developers reported the challenge of maintaining multiple packages, dependency hell, and the increase of duplicated packages.
As a way to cope with these challenges, they suggested grouping trivial packages. This could lead to a reduction in the number of dependencies by approximately 13\%.
\citet{Chowdhury2021TSE} empirically studied the project usage and ecosystem usage of trivial \npm packages. They reported that removing a trivial package can impact approximately 29\% of the ecosystem. They also revealed that trivial packages are being actively used in central \js files of software projects.

\subsection{Abandoned and unmaintained dependencies}

Proper dependency management does not suffice to consider technical factors only. Human factors are equally important.
Relying on reusable OSS packages all too often ignores the considerable effort required by package maintainers to keep them up to date and fixing reported issues in a timely manner~\cite{Wermke2023}. By depending on reusable packages, one implicitly trusts their associated OSS community. This can be problematic if this community --which is often driven by volunteers-- is not sufficiently active or responsive, is too small (e.g., packages maintained by a single developer), or if packages become unmaintained due to maintainer abandonment~\cite{Avelino2019empirical,Miller2023}.
\citet{Champion2021SANER} studied the \emph{underproduction} of software projects, where the
 supply of labour for maintaining them is too small to satisfy the demand of the project users.
 Analysing 21K+ packages with 461K+ bugs in the Debian distribution they proposed a method to identify underproductive software packages.

\citet{Miller2025ICSE} quantitatively observed that abandonment is common among widely-used \npm libraries, 
that hundreds of thousands of downstream projects were directly exposed to this abandonment,
and that most of these exposed projects never remove or replace an abandoned dependency. Factors that make it more likely to remove abandoned dependencies are more mature project government and dependency management practices, as well as the explicit public announcement of  abandoned packages.

Abandoned, and therefore unmaintained, libraries can be a source of highly impactful security threats. 
An incident in March 2024 was the compromised
\textsf{XZ-Utils} software compression package for Linux distributions. 
Its original well-intentioned maintainer who was no longer able to fully maintain the package.
After gaining this maintainer's trust during a period of two years, a malicious attacker took over its maintenance, and introduced a backdoor to authorise remote code execution on affected systems.
A similar situation happened in November 2018 for the \textsf{event-stream} \npm package, whose maintenance was unknowingly handed over to a malicious developer who subsequently modified the package to include code for stealing crypto-coins \cite{eventstream2018}.

These incidents underscore the importance of rigorous vetting processes to reduce the risk of social engineering that can compromise software integrity.
To avoid such incidents, it is of crucial importance to maintain a healthy and sustainable  community that is able to attract and retain motivated contributors \cite{Constantinou2017ISSE}, and to ensure that they have the necessary financial and computing resources to maintain their code. 
In case packages get abandoned, the ecosystem should rely on community package maintenance organizations (CPMOs), consisting of volunteers that steward and maintain abandoned packages~\cite{Zimmermann2023EMSE}.

\subsection{Incompatible licenses}
\label{sec:licence}

A final challenge stems from the incompatibility of OSS licenses that determine the terms and conditions to use or modify reusable libraries within one's own software. A plethora of licenses exist\footnote{The SPDX open standard for SBOM supports 600+ licenses, see \url{https://spdx.org/licenses/}}, making it increasingly challenging for developers to select an appropriate license for their projects and to ensure that they are complying with the terms of those licenses. 
Given that the use of incompatible licenses can lead to legal disputes, it is essential to ensure license compatibility when reusing OSS packages, and it is a frequent reason for migrating to an alternative library~\cite{He2021ESEC}.

\citet{XuLicense2023} conducted an empirical study of license incompatibilities and their remediation practices in the PyPI ecosystem for Python. They found that 7\% of the package releases have license incompatibilities and 61\% of them are due to transitive dependencies. They also identified five remediation strategies, including migrating to another library, removing the dependency, pinning versions or changing the license of the dependent package. Inspired by their findings, they proposed Silence, an approach to recommend license incompatibility remediations with minimal costs.
\citet{Wu2024MSR} conducted an empirical study of license usage, incompatibility and evolution in 33M+ packages across five package registries (\maven, \npm, \pypi, \rubygems and \cargo), observing both similarities and differences in license usage across the five registries.

\emph{Tool support.} In order to detect and resolve incompatible licenses, automated tools are needed. \citet{Xu2023TOSEM} proposed \textsf{LiDetector}, a learning-based tool to detect license incompatibilities. An empirical evaluation of 1,846 projects revealed that $>72\%$ of the projects suffer from license incompatibility, including popular ones such as the MIT License and the Apache License.
In a follow-up work~\cite{Xu2023ISSTA}, they presented \textsf{LiResolver}, a tool to resolve license incompatibility issues for OSS.

\section{Conclusion}

The reliance on reusable OSS libraries distributed through package registries continues to increase, and the size of their dependency networks follows suit.
Based on recent empirical research we compiled a catalogue of frequently studied dependency-related issues that these registries suffer from.
This overview of dependency challenges can be used as a starting point for researchers and practitioners that would like to delve deeper into the empirical research on OSS package dependency networks.

The extent to which they package registries suffer from dependency issues, and the way they deal with them, is highly ecosystem-dependent. There is no single ``one fits all'' solution given the diversity of targeted programming languages and communities involved in these ecosystems~\cite{Bogart2021TOSEM}.
Versioning policies such as semantic versioning have seen an increase in adoption. The same holds for the indispensable suite of tools to detect and fix dependency issues of various nature.
The use of appropriate policies and tools should be promoted further, and tools need to be improved and completed on a continuous basis (e.g., by including dynamic analysis techniques) to cope with the rapidly evolving OSS landscape.

Most of the empirical research focused on \npm and \maven, and to a lesser extent on \pypi. This is not surprising, since they come with the largest package registries, 
providing reusable libraries for \js, Java and Python respectively, the top three of most popular programming languages.\footnote{\url{https://spectrum.ieee.org/the-top-programming-languages-2024} [consulted on 1 September 2024]}
It is more surprising that some large package registries and popular programming languages have not received the  attention they deserve.
This is the case for \nuget, for instance, despite being the fourth-largest package registry (with 539K packages) and despite the popularity of C\# and the .NET platform.

With the increasing reuse of OSS packages comes an increasing reliance on software supply chains, hence research on software supply chain security and SBOMs is on the rise.
Security issues in package registries can also propagate well beyond the boundaries of the package registry, affecting other  ecosystems with an even larger attack surface. As an example, the \gha workflow ecosystem (created in 2018)  has been shown to suffer from vulnerabilities due to its dependence on reusable Actions implemented as \js components depending on \npm packages~\cite{OnsoriDelicheh2024MSR}.

\begin{acknowledgments}
We thank Stefano Zacchiroli, Jens Dietrich and Pol Dell'Aiera for their constructive feedback on this manuscript.
This work is supported by the Fonds de la Recherche Scientifique – FNRS under grant numbers J.0147.24, T.0149.22, and F.4515.23, as well as Action de Recherche Concertée ARC-21/25 UMONS3 financed by the Ministère de la Communauté française - Direction générale de l’Enseignement non obligatoire et de la Recherche scientifique.
\end{acknowledgments}


\end{document}